\documentclass{ETHpaper}

\usepackage[utf8]{inputenc}

\usepackage{amsmath}
\usepackage{xcolor}
\usepackage[capitalise]{cleveref}
\usepackage{tikz}
\usetikzlibrary{calc}
\usetikzlibrary{decorations.markings}
\usetikzlibrary{arrows, arrows.meta}
\usetikzlibrary{positioning}
\usepackage{pgfplots}
\pgfplotsset{compat=newest}
\usepgfplotslibrary{groupplots}
\usepackage{nicefrac}
\usepackage{url}
\usepackage{tabularx}
\usepackage{booktabs}
\usepackage{enumitem}
\usepackage{multirow}
\usepackage{makecell}
\usepackage{dcolumn}
\usepackage{url}
\usepackage{placeins}

\definecolor{colorSG}{RGB}{168,50,45}
\definecolor{customY}{HTML}{FBB13C}
\definecolor{customG}{HTML}{218380}
\definecolor{customT}{HTML}{73D2DE}
\definecolor{customW}{HTML}{FCFCFF}
\definecolor{customB}{HTML}{2E5EAA}
\definecolor{customP}{HTML}{5B4E77}
\definecolor{gitRedFaded}{HTML}{ffeef0}
\definecolor{gitGreenFaded}{HTML}{e6ffed}
\definecolor{gitRed}{HTML}{ffdce0}
\definecolor{gitGreen}{HTML}{cdffd8}
\definecolor{gitRedFull}{HTML}{cb2431}
\definecolor{gitGreenFull}{HTML}{2cbe4e}

\begin{document}

\title{Big Data = Big Insights? Operationalising Brooks' Law in a Massive GitHub Data Set}

\author{
\begin{tabularx}{\textwidth}{CC}
Christoph Gote$^{*}$ & Pavlin Mavrodiev$^{*}$\\
\normalfont\itshape cgote@ethz.ch & \normalfont\itshape pmavrodiev@ethz.ch\\[5mm]
Frank Schweitzer$^{*}$ & Ingo Scholtes$^{\dagger,\ddagger}$\\
\normalfont\itshape fschweitzer@ethz.ch & \normalfont\itshape ingo.scholtes@uni-wuerzburg.de\\
\end{tabularx}\\[5mm]
% \vspace*{10mm}
\normalfont\footnotesize\itshape
$^{*}$Chair of Systems Design, ETH Zurich, Weinbergstrasse 56/58, 8092 Zurich, Switzerland\\
$^{\dagger}$Chair of Computer Science XV - Machine Learning for Complex Networks, Julius-Maximilians-Universität Würzburg,\\Friedrich-Bergius-Ring 30, 97076 Würzburg, Germany\\
$^{\ddagger}$Data Analytics Group, University of Zurich, Binzmühlestrasse 14, 8050 Zurich, Switzerland
}

\authoralternative{Gote, et al.}
\www{\url{http://www.sg.ethz.ch}}
\reference{Version of: \today}

\date{May 2021}

\maketitle

\begin{abstract}
Massive data from software repositories and collaboration tools are widely used to study social aspects in software development.
One question that several recent works have addressed is how a software project's size and structure influence team productivity, a question famously considered in \emph{Brooks' law}.
Recent studies using massive repository data suggest that developers in larger teams tend to be less productive than smaller teams.
Despite using similar methods and data, other studies argue for a positive linear or even super-linear relationship between team size and productivity, thus contesting the view of software economics that software projects are \emph{diseconomies of scale}.

In our work, we study challenges that can explain the disagreement between recent studies of developer productivity in massive repository data.
We further provide, to the best of our knowledge, the largest, curated corpus of \textit{GitHub} projects tailored to investigate the influence of team size and collaboration patterns on individual and collective productivity.
Our work contributes to the ongoing discussion on the choice of productivity metrics in the operationalisation of hypotheses about determinants of successful software projects.
It further highlights general pitfalls in big data analysis and shows that the use of bigger data sets does not automatically lead to more reliable insights.
\end{abstract}

\section{Introduction}

Empirical research across disciplines is nowadays driven by the availability of big data and methods to process and analyse them efficiently.
In empirical software engineering, massive data from software repositories and online collaboration tools are widely used to investigate social and human aspects in software development.
This intersects with \emph{computational social science}, which uses big data to test hypotheses about individual and collective human behaviour originally developed in sociology, social psychology, or organisational theory.
Data-driven studies of developer productivity in large software projects are an exemplary case of how research in empirical software engineering can advance computational social science.
The question of how factors like, e.g., team size, influence the productivity of team members was already addressed by Maximilien Ringelmann~\cite{Ringelmann1913} in 1913.
In social psychology, his finding that individual productivity tends to linearly decrease with team size is known as the \emph{Ringelmann effect}.
In software project management, a similar observation is famously paraphrased as \emph{Brooks' law} \cite{brooks1975mythical}.
Here, the anecdote that ``adding manpower to a late project makes it later'' captures that the overhead associated with growing team sizes can reduce team efficiency.
Studies of collaborative software projects found evidence for a strong Ringelmann effect for different team sizes, programming languages, and development phases \cite{scholtes2016aristotle,gote2019analysing,blackburn1996improving,maxwell1996software}.
Other studies, however, found a positive linear or even super-linear relationship between the size of a team and the productivity of its members \cite{sornette2014much,muric2019collaboration,Maillart2019}.

The fact that different works studying the same research question yield qualitatively different results, despite applying similar methods to data from similar or even identical sources, should concern us.
Referring to the massive number of projects, commits, or developers covered in their studies, authors often corroborate their findings by the size of the data used to obtain them, thus implying that the analysis of bigger data automatically yields more reliable insights.
This points to an important general issue relevant for empirical research beyond software engineering:
Apart from advantages in terms of coverage, resolution, or statistical confidence, the use of big data also introduces new \emph{threats} for the validity of results.
To address this issue, in this work, we explore \emph{four challenges in the analysis of big data}. 
We study those challenges in a massive GitHub data set and argue that they are likely to explain conflicting results on the Ringelmann effect that were reported in recent works.

A first challenge is the \emph{\bfseries quality of big data} that, rather than being carefully collected and curated to address a specific research problem, are often incidentally generated as ``digital exhaust'' of large online platforms.
In empirical software engineering, this holds for massive data on software repositories harvested from online platforms like, e.g., \textit{GitHub} or \textit{SourceForge}.
While massive repository data promise insights into universals of collaborative software development, they are known to suffer from various quality issues.
These originate, e.g., from the inclusion of repositories that do not relate to software projects, projects whose development history is only partially represented in the data, or ambiguities that hinder the reliable identification of developers \cite{bird2009promises,kalliamvakou2014promises,kalliamvakou2016depth,gote2021gambit}.
This poses particular problems for studies addressing the effect of team size on developer productivity, which require reliable data on \emph{collaborative} projects that provide a complete picture of development actions attributable to individual team members.

A second challenge is \emph{\bfseries population validity}, which determines whether findings obtained in a given data sample can be extrapolated to a larger population.
On the one hand, for reasons of computational efficiency, researchers often base their results on a subset of the observations available in massive data, which can introduce biases that question population validity.
On the other hand, we can not necessarily avoid such biases by using all available data since population validity depends on the population for which we want to answer a given research question.
In massive repository data, using full information on all \textit{GitHub} projects may be justified if we want to answer a question about the population of \textit{GitHub} projects.
However, if we use data on \textit{GitHub} to obtain generalisable findings on how the size of software development teams affects the productivity of team members, we must carefully select projects to avoid biased samples in which either small or large teams are overrepresented.

A third challenge is \emph{\bfseries construct validity}, which includes the issue that rich and big data provide various options to operationalise research questions or hypotheses.
Whether or not the specific operationalisation chosen by a study is valid to address a research question is an important issue that influences the validity of results.
In the context of developer productivity, data on software repositories is an example of high-dimensional and time-resolved data.
In such data, productivity can be measured in various ways, and analyses can be applied for different levels of temporal aggregation, which is likely to affect the results.

Finally, \emph{\bfseries omitted-variable bias} is a fourth challenge that limits the reliability of findings if relevant variables are excluded from an analysis.
Technically, this is a general challenge that is not \emph{due} to the characteristics of big data. 
We nevertheless consider it in our work because high-dimensional data are likely to contain variables that can be used to address this issue.
In the context of Brooks' law, we can think of multiple explanations for an observed relationship between, e.g., the size of a team and the productivity of its members.
One explanation could be a \emph{causal} mechanism by which growing team size influences developer productivity, e.g., by reducing or increasing the motivation of team members.
An alternative explanation could be an additional variable related to the size of a team \emph{and} developers' productivity, such as e.g., the collaboration structure of a team.
A lack of control for such variables not only introduces biases in the inference of the actual relationship between variables of interest. 
It can also lead to the identification of spurious cause-effect relationships that negatively influence decision-making.

The four challenges summarised above question both the \emph{internal and external validity} \cite{campbell2015experimental} of empirical research in software engineering, which can explain why works studying the same question in the same data arrive at different conclusions.
Focusing on the operationalisation of Brooks' law, in this work, we show how to address them in massive \textit{GitHub} data.
Our contributions are as follows:
\begin{itemize}
\item[\raisebox{.05em}{\tiny\faPlay}] To address the challenges of \emph{\bfseries data quality} and \emph{\bfseries population validity}, we create a large, curated data corpus on Open Source Software (OSS) projects that facilitates the study of the influence of team size on both individual and collective productivity.
The projects included in this corpus are \emph{systematically} chosen based on (i) transparent filtering criteria that avoid common perils in GitHub mining \cite{kalliamvakou2014promises} and (ii) a stratified sampling that supports unbiased analyses of the impact of team size on developer productivity.
We make both our corpus and the pipeline to filter, sample, and process data based on \texttt{GHTorrent} \cite{gousios13ghtorrent}, a database freely available for researchers.
\item[\raisebox{.05em}{\tiny\faPlay}] To address \emph{\bfseries construct validity}, we systematically compare metrics for developer productivity in the data corpus created above.
Acknowledging that productivity is a multi-dimensional phenomenon, we select a set of eight code- and commit-based productivity measures. We study their cross-correlation to answer which of the measures are likely to be interchangeable and which capture independent dimensions of productivity.
\item[\raisebox{.05em}{\tiny\faPlay}] Addressing \emph{\bfseries omitted-variable bias}, we finally study to what extent changes in productivity can be causally explained by the collaboration structure of projects rather than team size.
Building on a recently developed method to construct time-evolving co-editing networks based on \texttt{git} repositories \cite{Gote2019git2net}, we investigate eight \emph{network metrics} whose choice is rooted in social capital theory.
We study the cross-correlation of those metrics to identify which of them capture independent dimensions.
\item[\raisebox{.05em}{\tiny\faPlay}] We apply our methods to study the Ringelmann effect in collaborative software development based on the corpus and methods developed above. We find a strong and significant negative relationship between team size and individual productivity that can be explained based on changes in the collaboration structure of software teams.
We further show that a failure to account for the challenges outlined above can lead to spurious results that suggest a \emph{positive} relationship.
\end{itemize}

In summary, we study challenges that can explain the disagreement between recent studies of developer productivity in massive repository data.
We further provide, to the best of our knowledge, the largest, curated corpus of \textit{GitHub} projects tailored to investigate the influence of team size and collaboration patterns on individual and collective productivity.
Our work contributes to the ongoing discussion on the choice of productivity metrics in the operationalisation of hypotheses about determinants of successful software projects.
It further highlights general pitfalls in big data analysis and shows that the use of bigger data sets does not automatically lead to more reliable insights.

\section{Systematic Construction of Data Corpus}\label{sec:data_corpus}

We first introduce a framework to select and mine projects from \textit{GitHub} that can be used to address the first two challenges of \emph{data quality} and \emph{population validity}.
We use it to systematically sample 201 OSS projects covering the entire range of team sizes on \textit{GitHub}.
We further extract time-stamped editing events that we use to investigate whether collaboration structures can explain team productivity.

\subsection{Data quality}\label{sec:challenge_of_selecting_software_repositories}

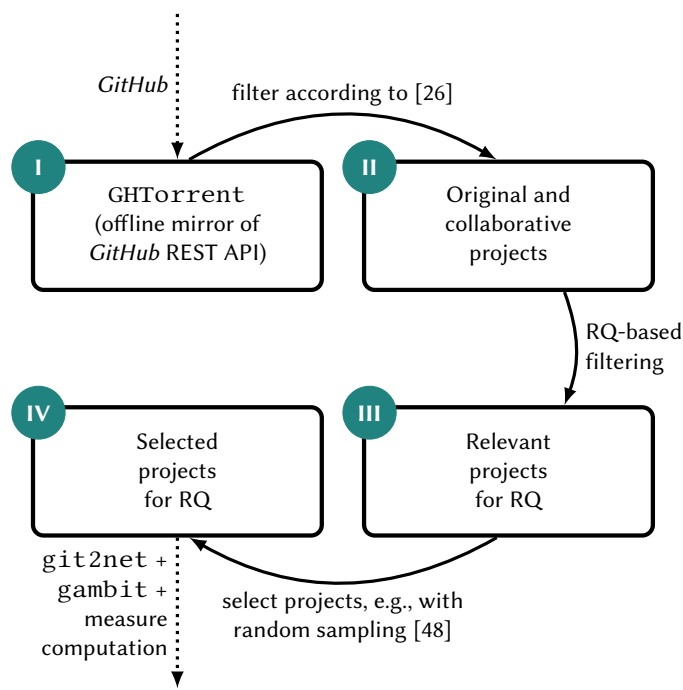
\begin{figure}[t!]
    \centering
    \begin{tikzpicture}\sffamily\small
        \tikzstyle{stepbox} = [node distance=5mm, align=center, draw=black, rounded corners, ultra thick, minimum width=38mm, minimum height=17mm]
    
    	\node[stepbox] (GHT) {\texttt{GHTorrent}\\ (offline mirror of\\ \textit{GitHub} REST API)};
    	\node[stepbox, right=of GHT] (SDP) {Original and\\collaborative\\projects};
    	\node[stepbox, node distance=15mm, below=of SDP] (FPR) {Relevant\\projects\\for RQ};
    	\node[stepbox, left=of FPR] (SPR) {Selected\\projects\\for RQ};
    	
    	\node[circle, white, draw=customG, fill=customG, inner sep=0, minimum width=7mm, xshift=1mm, yshift=-1mm] at (GHT.north west) {\textbf{I}};
    	\node[circle, white, draw=customG, fill=customG, inner sep=0, minimum width=7mm, xshift=1mm, yshift=-1mm] at (SDP.north west) {\textbf{II}};
    	\node[circle, white, draw=customG, fill=customG, inner sep=0, minimum width=7mm, xshift=1mm, yshift=-1mm] at (FPR.north west) {\textbf{III}};
    	\node[circle, white, draw=customG, fill=customG, inner sep=0, minimum width=7mm, xshift=1mm, yshift=-1mm] at (SPR.north west) {\textbf{IV}};
    	
    	\path[-latex, very thick] (GHT.80) edge[bend left=30] node[align=center, anchor=south] {filter according to \cite{kalliamvakou2014promises}} (SDP.100);
    	\path[-latex, very thick, rounded corners] (SDP.310) edge[bend left=20] node[align=left, anchor=west] {RQ-based\\ filtering} (FPR.50);
    	\path[-latex, very thick] (FPR.260) edge[bend left=30] node[align=center, anchor=north] {select projects, e.g., with \\ random sampling \cite{seawright2008case}} (SPR.280);
    	
    	\draw[latex-, dotted, very thick] (GHT.90) -- node[pos=.5, left] {\textit{GitHub}}  ++ (0,2);
    	\draw[-latex, dotted, very thick] (SPR.270) -- node[pos=.45, left, align=right] {\texttt{git2net}~+ \\ \texttt{gambit}~+ \\ measure \\ computation} ++ (0,-2);
	\end{tikzpicture}
	
    \caption{Pipeline to select and sample collaborative software development projects from \textit{GitHub} to address a given research question (RQ).}
    \label{fig:project_selection_pipeline}
\end{figure}

To select suitable projects from \textit{GitHub}, we propose the project selection and sampling pipeline shown in \Cref{fig:project_selection_pipeline}.
As a first step, we need to gain access to the information required to apply our selection criteria.
Because \textit{GitHub}'s REST API is rate-limited to 5,000 requests per hour\footnote{\label{fn:june2020} as of June 2020}, retrieving the metadata of more than 100 million repositories hosted on \textit{GitHub} \cite{warner2018repositories} becomes untenable.
We, therefore, use the database made available by the \texttt{GHTorrent} project \cite{gousios13ghtorrent}, which has crawled most of \textit{GitHub}'s REST API using donated API keys.
We use the latest\footnotemark[1] available dump from June 2019 that contains data on a total of more than 125 million repositories.

As a second step, we determine which projects in the \texttt{GHTorrent} database are suitable to study \textit{collaborative OSS development}.
It has already been reported by the authors of \cite{kalliamvakou2014promises} that the majority of projects on \textit{GitHub} are either personal, inactive, or very small and should be excluded when analysing collaborative software development.
We adopt the filtering criteria proposed by the authors of \cite{kalliamvakou2014promises}, namely excluding repositories with a single developer, fewer than 50 commits, or a span of fewer than 100 days between the first and last commit in the repository.
The resulting data set reduces to around 4.5 million OSS projects, i.e., 3.6\% of all \texttt{GHTorrent} repositories.
In other words, following \cite{kalliamvakou2014promises}, at least 96.4\% of the repositories in the latest \textit{GHTorrent} database are not suitable for studying collaborative software development.

We further improve on these filtering criteria in the following two ways.
First, \textit{GitHub} is frequently used for applications not related to software development, such as free file storage or web hosting \cite{kalliamvakou2014promises}.
Our own analysis revealed a substantial number of popular repositories\footnote{popular judged by a high count of forks, commits, developers, or stars} not representative of collaborative software development, e.g., tutorials on \textit{git}, code snippet repositories, or \textit{git}-based ``clocks'', which are updated with a new commit every second.
Excluding these repositories is crucial to avoid misleading results.

Second, as a result of the functionality to \textit{fork} any public repository, \textit{GitHub} contains a substantial amount of repositories that are in large parts exact copies of other projects.
We drop all repositories that are designated as \textit{forks} to avoid biases from analysing the same commit history multiple times.

After applying these two additional filtering steps, we retain a list of around 1.8 million original and collaborative repositories, which is only 1.4\% of all repositories in \texttt{GHTorrent}.
This underlines the importance of proper data selection when analysing collaborative software development on \textit{GitHub}.

\subsection{Population validity}\label{sec:D1}

Besides projects that are original and collaborative, our study on team productivity requires projects that fulfil additional conditions regarding their (i) activity, (ii) size, and (iii) purpose, i.e., collaborative projects for developing software (step 3 in \Cref{fig:project_selection_pipeline}).

\paragraph{Project activity}
To avoid issues that could arise from mixing active projects with those where development has ceded long in the past, we focus on projects that are actively developed at the time of our study.
We regard a project as active if the last commit was made after May 2020.
To ensure this, we take a two-fold approach: we first select projects from \texttt{GHTorrent} that have a recorded commit activity after May 2019. 
In a subsequent step, we then use the \textit{GitHub} REST API to filter those projects that additionally have a recorded commit after May 2020.

\paragraph{Project size}
To facilitate an unbiased sampling of projects based on team size, we first need to determine the size of a development team.
For OSS projects without formal team memberships, this is a challenging task.
The authors of \cite{scholtes2016aristotle} found that the probability of making future contributions to an OSS project drops below 10\% after an inactivity of approx. 42 weeks.
Based on this finding, we compute the size of an OSS development team at time $t$ by counting all developers who committed within a moving time window of 294 days, i.e., between $t - 294$ days and $t$.
Thus, to compute a team size, projects need to have existed for at least 294 days, increasing the requirements beyond the 100 days considered by the authors of \cite{kalliamvakou2014promises}.

\begin{table}[t!]
    \centering
    \caption[Number of projects with different team size ranges]{Number of projects with different team size ranges. All given team size ranges include the outer values. Team sizes are computed based on the data available in \texttt{GHTorrent}.}
    \sffamily\small
    \begin{tabularx}{.22\textwidth}{r@{\hspace{1mm}--\hspace{1mm}}Xrr}
        \toprule
        \multicolumn{2}{c}{\multirow{2}{*}{\makecell{\\[-1.4mm]\textbf{Team}\\ \textbf{Size}}}} & \multicolumn{2}{c}{\textbf{Projects}}\\
        \cmidrule(lr){3-4}
        \multicolumn{2}{c}{} & \multicolumn{1}{c}{\textbf{Tot.}}  & \multicolumn{1}{c}{\textbf{Sel.}} \\
        \midrule
        2 & 4 & 95,763 & 19\\
		5 & 8 & 33,027 & 27\\
		9 & 15 & 17,221 & 26\\
		16 & 30 & 10,027 & 27\\
		31 & 58 & 3,499 & 27\\
		\bottomrule
    \end{tabularx}\hspace{5mm}
	\begin{tabularx}{.22\textwidth}{r@{\hspace{1mm}--\hspace{1mm}}Xrr}
        \toprule
        \multicolumn{2}{c}{\multirow{2}{*}{\makecell{\\[-1.4mm]\textbf{Team}\\ \textbf{Size}}}} & \multicolumn{2}{c}{\textbf{Projects}}\\
        \cmidrule(lr){3-4}
        \multicolumn{2}{c}{} & \multicolumn{1}{c}{\textbf{Tot.}}  & \multicolumn{1}{c}{\textbf{Sel.}} \\
        \midrule
		59 & 115 & 1,476 & 25\\
		116 & 226 & 665 & 16\\
		227 & 443 & 231 & 13\\
		444 & 871 & 102 & 15\\
		872 & 1,711 & 28 & 6\\
		\bottomrule
    \end{tabularx}
    \label{tab:log_strata}
\end{table}

\Cref{tab:log_strata} shows the number of projects for different team sizes, where team size is computed for the latest available 294-day time window.
The project counts are reported for ten $\log_2$-spaced strata, which yields a distribution where the team size roughly doubles for each consecutive stratum.
The resulting distribution is right-skewed, where the vast majority of projects have small team sizes.
A uniform sample from the complete set of projects would thus primarily select small projects, which would fail to cover a broad spectrum of team sizes.
To remedy this, we sample 28 projects from each stratum, where 28 is the size of the stratum with the fewest projects.

While sampling, we ensure that all sampled projects are software development projects and continue to be actively developed at the time of mining.
We further remove duplicate projects that originate from manual clones of other repositories.
We achieve this by applying the following selection criteria:

\paragraph{Project purpose}
To identify a project's purpose, we query the \textit{GitHub} REST API to obtain the most recent information on all considered projects.
We first ensure that all sampled projects are software development projects and continue to be actively developed at the time of mining.
We consider a project as a software development project if at least 75\% of the code in the repository is written in the 17 programming languages supported by the code analysis tool \texttt{lizard} \cite{yin2020lizard}.
In total, the languages supported by \texttt{lizard} account for over 85\% of the code submitted to \textit{GitHub} \cite{beuke2020githut}.

\paragraph{Deduplication}
Finally, our set of projects still contains duplicate repositories that originate from manual clones pushed to a different repository rather than using the fork mechanic recorded in the \texttt{GHTorrent} data.
Removing these clones is an important challenge when selecting repositories for analysis, and independent data sets listing duplicate repositories have been developed \citep{spinellis2020dataset}.
Unfortunately, these data sets were not yet available at the time of our analysis.
Therefore, we manually removed the clones, retaining the original project that was cloned.

The two selection criteria require us to query the \textit{GitHub} REST API or perform manual filtering, respectively.
Due to the API's rate limit, this means that neither can be performed at large scale before sampling projects.
Instead, they need to be performed during the sampling process.
We treat all strata equally and apply the additional selection criteria to the sampled 28 projects from each stratum.
As shown in the final column of \Cref{tab:log_strata}, this yields between 6 and 27 projects for each of the ten strata, resulting in a total of 201 projects with a total of more than 100,000 developers and over 3 million commits (step 4 in \Cref{fig:project_selection_pipeline}).
Overall, we obtain relatively similar project counts for all strata, except for the strata with largest and smallest team sizes.

\subsection{Mining co-editing networks from git repositories}\label{sec:retrieving-interactions-and-measure-computation}

We mine all edits and co-edits for the full history of the 201 projects using the Open Source Python tool \texttt{git2net} \cite{Gote2019git2net}.
Besides co-editing relations, we also extract both commit- and code-based productivity measures.
To this end, we apply \texttt{lizard} \cite{yin2020lizard} and an optimised version of \texttt{multimetric} \cite{weihmann2020multimetric} to the source code before and after each change.
Obtaining highly granular information on the development process of over 200 OSS projects requires substantial computational resources, in our case, over 1 million CPU-hours.
Therefore, we perform all computations on 256 compute cores within a time frame of over six months on the ETH Zurich scientific compute cluster \textit{Euler}.

An additional challenge in the analysis of \textit{git} repositories is the need to disambiguate commit authors.
This step is necessary as developers can make contributions using different credentials, e.g., due to spelling errors in usernames or the use of nicknames.
Considering different aliases as different users would lead us to overestimate the team size and underestimate the productivity of developers with multiple aliases.
We thus use the recently proposed tool \texttt{gambit} \cite{gote2021gambit} to disambiguate all developers in all repositories.

Upon manual inspection, we found that some projects contain very large commits originating from code imports or automated code refactoring tools.
Such, mostly automated, commits are not representative for the coordination requirements between developers.
However, due to their size, they could lead to a bias our subsequent analysis.
Therefore, as a final data cleaning step, we drop outliers by excluding all commits outside the $2.5$th and $97.5$th percentile regarding their total Levenshtein distance.

A complete list of projects as well as anonymised raw data of all projects considered in our analysis is archived on \url{zenodo.org}\footnote{https://doi.org/10.5281/zenodo.5294964}.

\section{Operationalising Productivity and Collaboration Structure}
\label{sec:operationalizing}

To study how team size affects the productivity of OSS projects, we need to operationalise (i) the size of OSS teams, and (ii) the productivity of OSS teams.
In addition, we aim to understand how coordination between different team members affects this relation.
Therefore, we need to also operationalise (iii) the collaboration structure of OSS teams.

We base our operationalisations of all three concepts on the edits and co-edits observed within non-overlapping 42-week time windows.
As discussed in \Cref{sec:D1}, the choice of 42 weeks is motivated by \citep{scholtes2016aristotle} who found that after this time, the probability of a developer making future contributions to a project is less than 10\%.
Ensuring that the time window is divisible by full weeks is essential to ensure that the weekly productivity patterns present on \textit{GitHub} \citep{github2020weeklypattern} do not bias our results.

In the next three sections, we will discuss each of the operationalisations in detail.

\subsection{Team size}
OSS projects utilise the principles of open collaboration to create new software.
This means that they rely on contributions of loosely coordinated participants, who differ significantly regarding the size of their contributions.
Contributors can further join and leave the team at any time.
Due to this method of collaboration, no organised ledgers listing the members of OSS teams exist.
This makes operationalising the size of such teams non-trivial.

The consensus of prior literature is that all individuals contributing to an OSS project should be considered as team members \citep{vasilescu2015gender}.
With this work, we study the production of code artefacts.
Therefore, we operationalise team size as the count of all individuals who contribute code to a project within a given time window.
This includes all developers adding, modifying, or removing code from the project's codebase.

\subsection{Productivity measures}\label{sec:productivity_measures}

\begin{table}
    \centering
    \caption[Productivity measures considered in this paper]{Productivity measures considered in this paper. All measures are evaluated over a time window of length $\Delta t$ and normalised by the team size (TS).}
    \sffamily\small
    \begin{tabular}{cll}
        \toprule
        \multirow{3}{*}{\rotatebox{90}{\makecell{Commit-\\Based}}} & \textbf{Comms} & commits $/\Delta t/ \text{TS}$ \\[.5mm]
        & \textbf{Events} & lines added, modified, or deleted $/\Delta t/ \text{TS}$\\[.5mm]
        & \textbf{LevD} & characters modified $/\Delta t/ \text{TS}$\\
        \midrule
        \multirow{5}{*}{\rotatebox{90}{\makecell{Code-\phantom{---}\\Based\phantom{---}}}} & \textbf{NLOC} & lines of code changed $/\Delta t/ \text{TS}$\\[.5mm]
        & \textbf{Tokens} & change in number of tokens $/\Delta t/ \text{TS}$\\[.5mm]
        & \textbf{Funcs} & change in the number of functions $/\Delta t/ \text{TS}$\\[.5mm]
        & \textbf{CycC} & change in cyclomatic complexity $/\Delta t/ \text{TS}$\\[.5mm]
        & \textbf{HalEff} & Halstead effort to make changes $/\Delta t/ \text{TS}$\\
        \bottomrule
    \end{tabular}
    \label{tab:productivity_measures}
\end{table}

\begin{figure*}
    \centering
    \includegraphics[width=\textwidth]{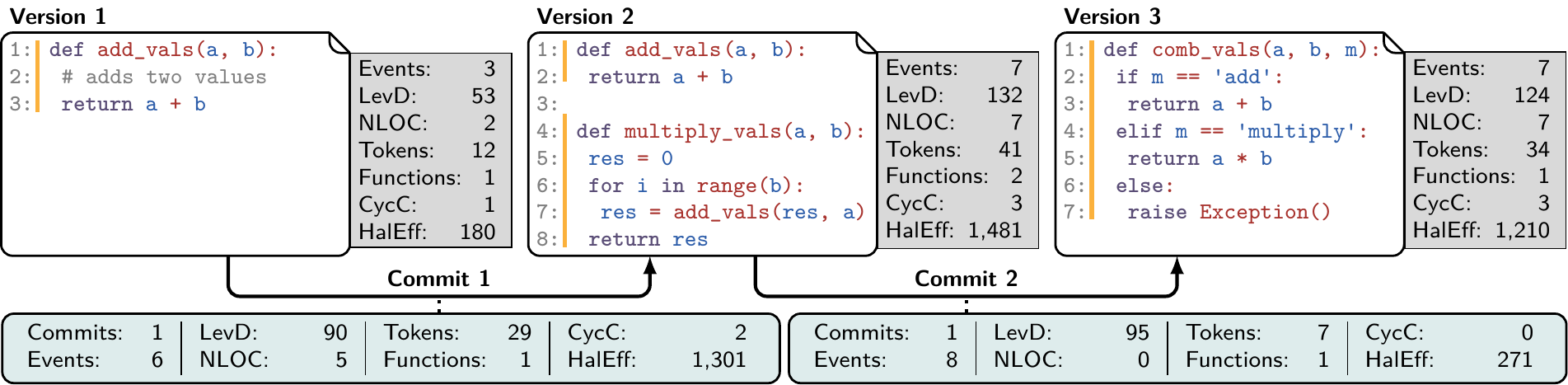}
    \caption{Productivity measures applied to three consecutive versions of an exemplary Python file. In the grey boxes, we report the productivity associated with creating each version of the file from scratch.
    In the green boxes, we show the productivity related to the changes observed between the versions.
    We assume that each new version was created in a single commit.
    All tokens are highlighted. Tokens that are operands and operators are printed in blue and red, respectively. All other tokens are printed in purple. Individual functions are indicated by yellow bars. For the computations in the example, we assume $\Delta t = \text{TS} = 1$.}
    \label{fig:Productivity_Visualisation}
\end{figure*}

Before discussing how we operationalise team productivity, we need to precisely define this term.
Productivity captures one aspect of the broader concept of \emph{team effectiveness}.
Here, team effectiveness is defined as (i) the productive output of the work group, (ii) the effectiveness of processes to maintain the team's capability in the future, and (iii) the satisfaction of group member's personal needs \citep[p. 323]{hackman1987design}.
With our study on team productivity, we focus on the first aspect.
Specifically, we assess the input-output relation considering the size of the code changes made by an OSS development team as a function of the team's size.

To operationalise team productivity, we need to define measures that allow us to capture the size of a change in a project's codebase.
For this, many different measures have been proposed in the literature.
Addressing \emph{construct validity}, we consider eight productivity measures and investigate the extent to which they provide independent information on the construct of productivity.
We further carefully investigate and assess how these measures interrelate.

We categorise our eight productivity measures as commit- or code-based measures \citep{oliveira2020code}.
Commit-based metrics rely solely on the size of the changes within a repository, e.g., the number of commits, the number of changed lines, or the number of modified characters.
Commit-based productivity measures require low computational effort and are independent of the programming language used in a repository.
However, by not assessing the content of a repository's code, they do not allow us to distinguish, e.g., between lines of code or comments that are added.
Therefore, we also consider code-based measures that take these aspects into account.
These include measures based on the number of modified lines of code (NLOC), the number of code tokens or functions, changes in McCabe's cyclomatic complexity \citep{mccabe1976complexity}, or the Halstead effort \citep{halstead1977elements}.
We provide a complete overview of the productivity measures considered in our study in \Cref{tab:productivity_measures}.
We compute productivity for each time window and normalise the productivity by the respective team size (TS).

To illustrate how different productivity measures can introduce the challenge of construct validity, consider the exemplary Python code shown in \Cref{fig:Productivity_Visualisation}.
We start with version one of a file that contains three lines of text with a total of 53 characters (whitespaces included).
Creating this file from scratch would require three \emph{line modification events}, i.e. three line-additions where lines 1 and 3 contain actual code, and line 2 contains a comment.
Therefore, the number of line modification events is three, while the lines of code (NLOC) is two.
In the example, we have highlighted all code tokens.
To compute the Halstead effort, we need to distinguish between tokens that are operands and operators.
These are highlighted in blue and red, respectively, whereas all other tokens are printed in purple.
In total, we have $12$ tokens.
With \texttt{a} and \texttt{b}, the code contains $\eta_2 = 2$ distinct operands that appear a total of $N_2 = 4$ times.
We further have $N_1 = 6$ operators that are all different from each other (i.e. $\eta_1 = 6$). 
The Halstead effort to create this file is thus defined as $E = (N_1 + N_2)\cdot\log_2(\eta_1 + \eta_2)\cdot\frac{\eta_1}{2}\cdot\frac{N_2}{\eta_2} = 180$.
Finally, the file only has one function without any branches resulting in $\text{Functions} = \text{CycC} = 1$.

With the first modification, we add a second function implementing the multiplication of two values.
The bottom-left green box in \Cref{fig:Productivity_Visualisation} shows the productivity of this change.
The code-based productivity measures are computed as the productivity difference to create the two consecutive versions of the file.
For the commit-based measures, the contents of the two files are compared directly.
With the second modification, we merge the two functions into one.
Despite the increase in characters, the token and function counts and the Halstead effort of version three are lower than those of version two.
If measures such as the number of functions or the cyclomatic complexity were to only increase, this would make the code difficult to maintain and prone to bugs \citep{stamelos2002code, baggen2012standardized}.
Therefore, we also consider contributions reducing code complexity, e.g., by consolidating functions or refactoring code, as productive.
We achieve this by computing the productivity of a modification as the absolute value of the productivity values to create the versions before and after the observed change.

This simple example shows that our eight productivity measures can yield considerably different results.
This prompts the question of which measure we should use as a target variable that we seek to explain through the set of features identified above.

We address this question in an exploratory study analysing the 201 OSS repositories in our corpus.
For this, we split time-series data into non-overlapping 42-week time windows.
We then compute our productivity measures and drop all time windows in which a team was inactive, i.e., for which we observe a productivity of zero, yielding a total of 1,188 observations.
We find that the distributions of all productivity measures are highly skewed.
Therefore, we log-transform all skewed measures such that the resulting distributions resemble a normal distribution.

\Cref{fig:corrplots}a shows the Pearson correlation between all productivity measures.
We find values larger than 0.9 between all productivity measures except for the number of commits and Halstead effort.
This suggests that the change in both characters and tokens is similar to the change in lines.
We further find that the number of functions and cyclomatic complexity are positively correlated, both changing with the number of lines.
With values between 0.7 and 0.8, correlations are considerably smaller for the number of commits and Halstead effort.
This indicates that commits differ considerably in terms of their size, i.e., with regard to the number of characters, lines, tokens or functions modified with the commit.
Halstead effort is unique among the considered productivity measures as, next to the total amount of code, it also considers the size of the vocabulary used.
Therefore, slower vocabulary growth compared to the total amount of code could explain the observed smaller correlation with other measures.

In conclusion, all considered productivity measures have different motivations.
Some analyse the source code at various levels of detail while others aggregate information at the level of lines or commits.
Despite those differences and the strong differences shown in the example in \Cref{fig:Productivity_Visualisation}, in our corpus of projects, and when computing average developer productivities across teams, all productivity measures are highly correlated.

\subsection{Collaboration networks of OSS teams}
\label{sec:colaboration_nets}

\begin{figure*}
    \centering
    \scalebox{1.05}{
    \begin{tikzpicture}\sffamily
        \node (n1) {\includegraphics[height=.25\columnwidth, trim=50 73 5 65, clip]{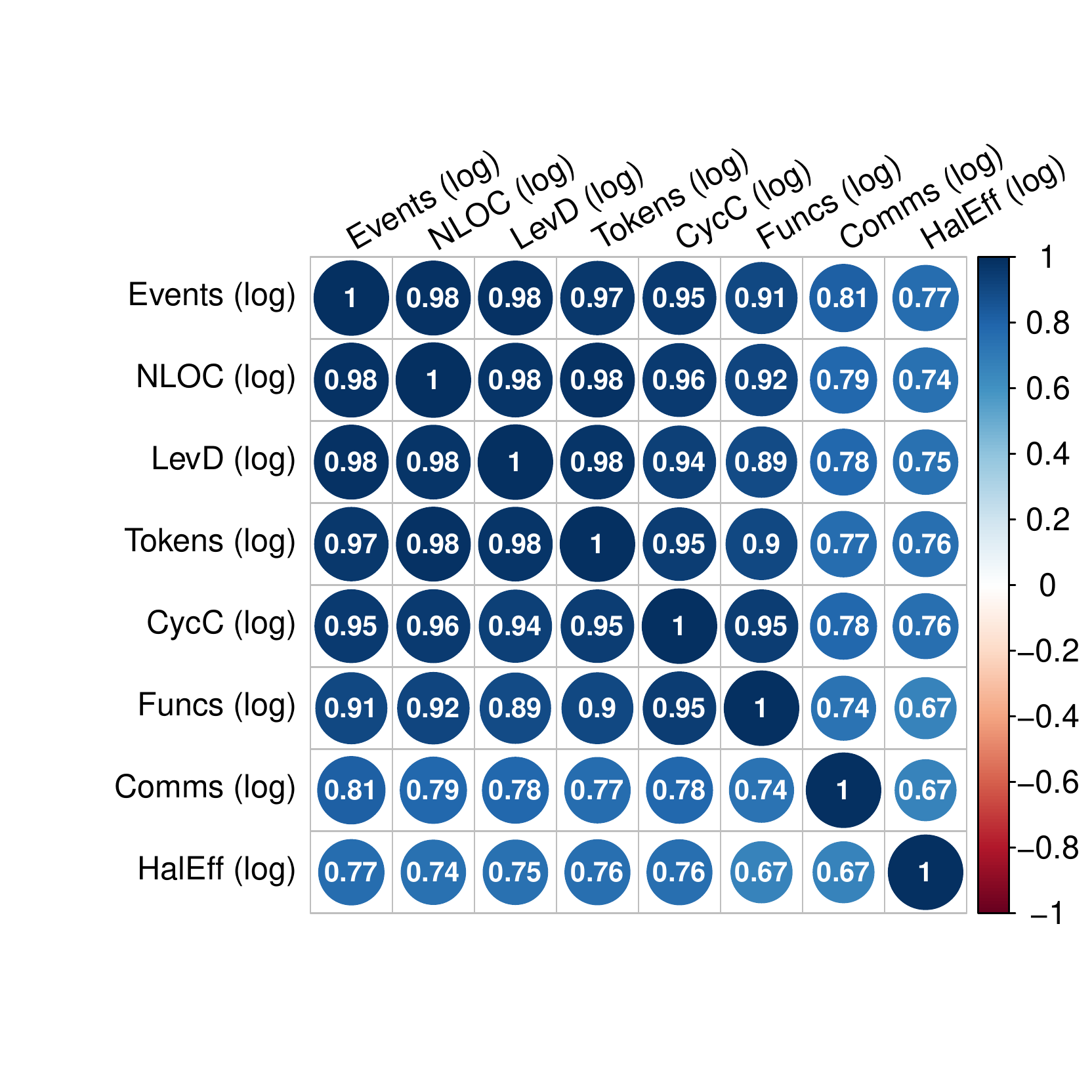}};
        
        \node[node distance=0, right=of n1,yshift=-.25mm] (n2) {\includegraphics[height=.2462\columnwidth, trim=52 71 5 60, clip]{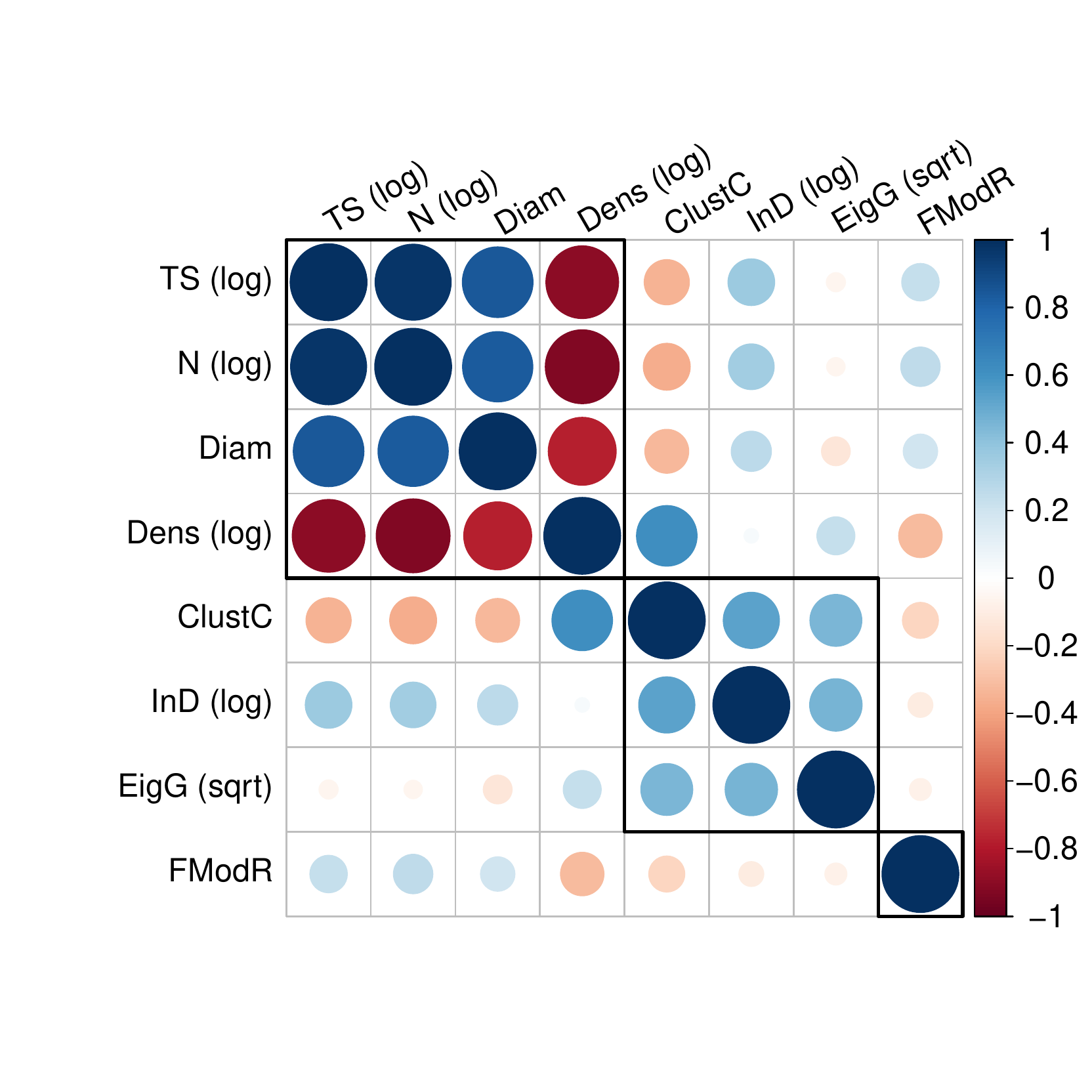}};
        
        \node[node distance=0, right=of n2,yshift=-.1mm] (n3) {\includegraphics[height=.2543\columnwidth, trim=52 65 5 55, clip]{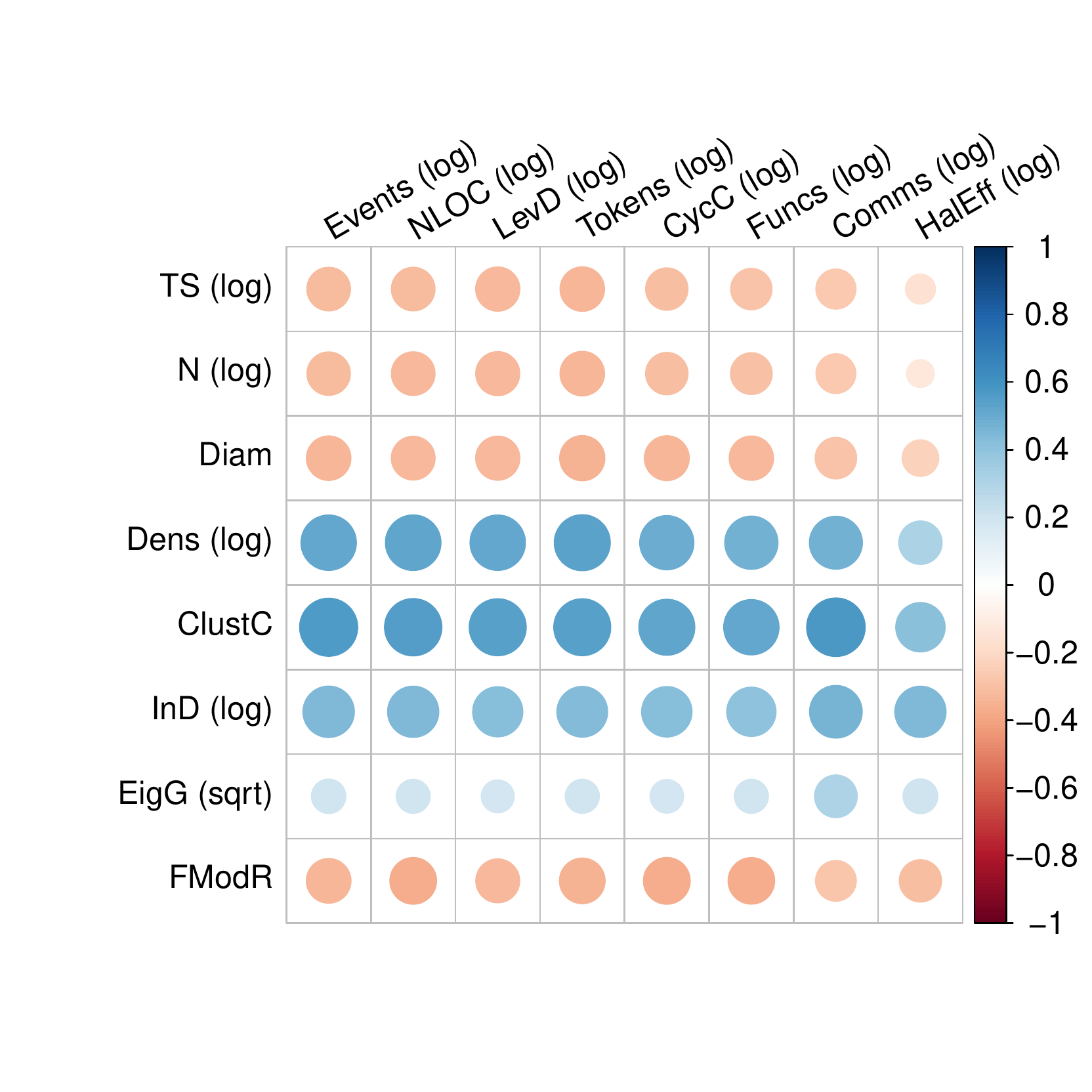}};
        \node[anchor=north west,inner sep=0,xshift=1mm, yshift=2mm] at ($(n1.north west |- n2.north west)$) {\textbf{A}};
        \node[anchor=north west,inner sep=0,xshift=1mm, yshift=2mm] at (n2.north west) {\textbf{B}};
        \node[anchor=north west,inner sep=0,xshift=1mm, yshift=2mm] at ($(n3.north west |- n2.north west)$) {\textbf{C}};
    \end{tikzpicture}
    }
    \caption{Results of exploratory study on team productivity and collaboration structure. a) Pearson correlation between the transformed productivity measures. b) Pearson correlation between the transformed network measures. Clusters between the measures are marked. c) Cross-correlation (Pearson) between the transformed network and productivity measures.}
    \label{fig:corrplots}
\end{figure*}
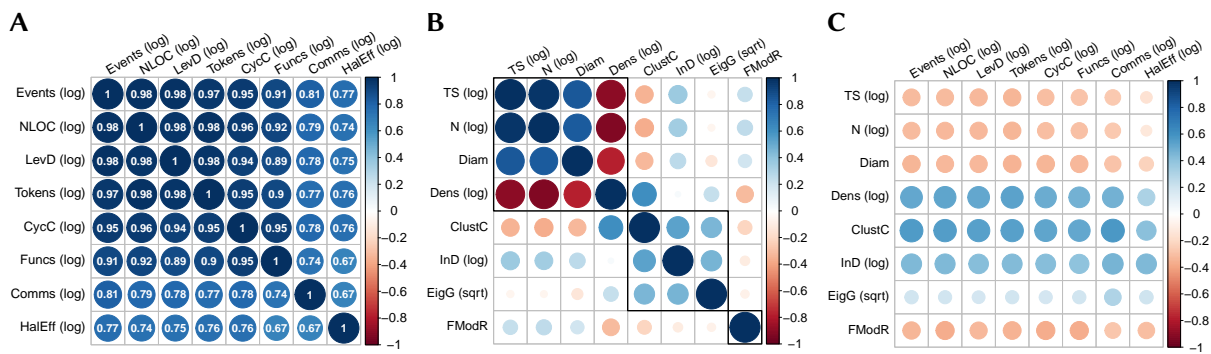

Addressing the challenges of \emph{omitted-variable bias}, we explore how the collaboration structure of teams might modulate team productivity.
In this way, we capture team characteristics beyond mere team size, highlighting additional variables that we need to control for when studying Brooks' law in rich data.
The inclusion of additional measures capturing collaboration structure was motivated by \citep{scholtes2016aristotle}, which found that a team's network structure affects the slope of the relation in their data.
Therefore, for this work, we consider network-based measures that fit this prior work.
In addition, we also include software-engineering-specific measures.

Specifically, to capture characteristics of different aspects of the collaboration structures of development teams, we use measures that we compute on the co-editing network constructed for our non-overlapping 42-week time windows.
In these co-editing networks, nodes represent different developers and edges $A \rightarrow B$ represent events where developer $B$ modifies a line of code last edited by $A$. 
The direction of the edge indicates the change of line ownership from $A$ to $B$.
Multiple co-editing events between two developers are represented as multi-edges between nodes.
Developers editing their own code are captured as self-loops.
In the following, we present eight measures that can be used as control variables to explain the relation between team productivity and team size.
For formal definitions,  we refer to \citep{newman2018networks}.

\paragraph{Number  of  nodes  (N)}
The number of nodes in the co-editing network counts all developers that actively edited code or whose code was edited.
The number of nodes is always greater than or equal to the team size.

\paragraph{Number  of  edges  (Edges)}
The number of edges counts the co-editing events within a time window.

\paragraph{Density  (Dens)}
The density captures the proportion of potential edges present in  
a network.
We compute the density based on the flattened network, in which multi-edges between two nodes are substituted by a single edge.

\paragraph{Diameter  (Diam)}
The network’s diameter is given by the length of the longest shortest path between any pair of nodes.

\paragraph{Clustering  Coefficient  (ClustC)}
A  node’s local clustering coefficient is computed as the fraction of pairs of neighbours that are connected by an edge.
The global clustering coefficient is obtained as the average local clustering coefficient across all nodes in the network.
Networks with small diameter and large clustering coefficient exhibit the so-called small-world property.
Links that connect different clusters in a network lead to low diameters, even for networks with many nodes.
The small-world property is directly related to navigability, knowledge transfer and social capital within social networks \citep{milgram1967small,granovetter1973strength}.

\paragraph{Mean   Indegree   (InD)}
In the flattened co-editing network, the indegree of a  node $i$ indicates the number of developers whose code has been edited by $i$.

\paragraph{Mean Foreign Modification Ratio (FModR)}
A recent work shows that the productivity of developers is significantly reduced if they edit code owned by other developers compared to editing their own code \citep{gote2019analysing}.
We account for this using the foreign modification ratio, which we compute as the fraction of all co-editing events where the developer edits code owned by another developer.
We obtain the number of all co-edits of a developer $i$ as $i$'s indegree, and the number of co-edits where $i$ edit foreign code as the count of all edges to $i$ that are not self-loops.
The mean foreign modification ratio of the team is obtained as the mean foreign modification ratio of all team members.

\paragraph{Eigengap (EigG)}
Finally, the eigengap, also referred to as spectral gap, of a network captures the efficiency of dynamical processes on the network.
Networks with larger eigengaps support fast spreading, diffusion and synchronisation, which can be interpreted as a proxy for the efficiency of information exchange and consensus schemes.
We compute the eigengap for the largest connected component of the network.

The definitions above enable us to capture the collaboration structure of software development teams in a multi-dimensional feature space.
Similar to the productivity measures, we find that the distributions of some network measures are highly skewed.
Therefore, we again apply logarithmic or square-root transformations.
We report the applied transformation for all measures throughout the remainder of this manuscript.

We next aim to select a minimum set of features that capture independent dimensions of collaboration networks.
For this, we study pair-wise correlations between all features, identify clusters of highly correlated features, and select one representative feature per cluster.
While we could instead use dimensionality reduction techniques like principal component analysis, our approach provides the advantage that it allows us to analyse interpretable network features rather than principal components.

\Cref{fig:corrplots}b shows the Pearson correlation between all pairs of network measures.
A first visible cluster in the upper-left quadrant contains team size, the number of nodes, and the network diameter, which all show a strong positive correlation.
In addition, network density is strongly negatively correlated with all three.
Thus, the relative number of co-editing interactions goes down for larger teams, leading to the distance between the two furthest team members in the co-editing network to increase.
The second cluster contains clustering coefficient, mean indegree, and eigengap, which are all positively correlated.
Thus, in teams where everyone interacts with many different team members we obtain a network structure in which information can spread more quickly throughout the team.
The third cluster contains only the mean foreign modification ratio, which quantifies how much other developers' code is edited within the team.
For all downstream analyses, we select team size (TS), mean indegree (InD) and the foreign modification ratio (FModR) from the set of network measures, i.e., we use one measure from each cluster.

We finally consider cross-correlations between the network and productivity measures shown in \Cref{fig:corrplots}c.
The results of this analysis confirm that all productivity measures exhibit very similar correlations to the network metrics.

Overall, with the results from our correlation studies, we confirm and extend the prior findings on the relations between social network measures for OSS projects \citep{allaho2015analyzing,teixeira2015lessons} and the relations between classical source code metrics \citep{henry1981relationships,woodfield1981study,mamun2017correlations,landman2014empirical} for a broader set of measures and in our novel and significantly more extensive corpus of projects designed to study the productivity of OSS development teams.

\section{Testing Brooks' Law in Massive GitHub Data}

\begin{figure*}
    \centering
    \includegraphics[width=\textwidth, trim=5 0 12 20, clip]{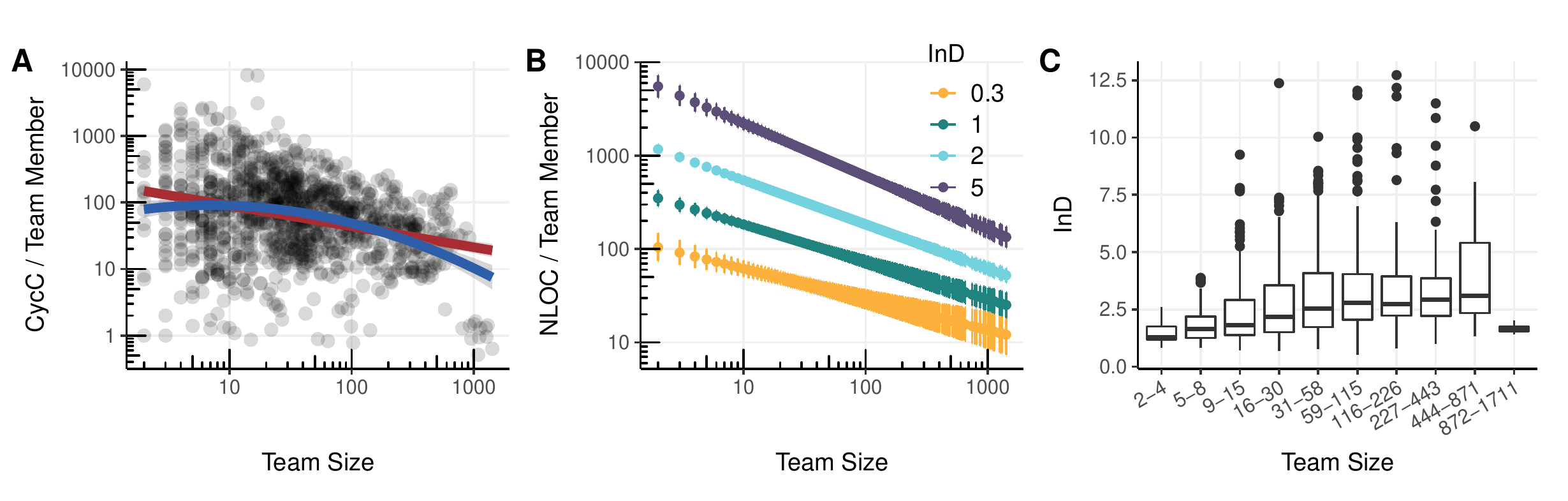}
    \caption{a) Productivity (CycC) per team member as a function of team size. A linear {\textcolor{colorSG}{\footnotesize\faSquare}} and quadratic model {\textcolor{customB}{\footnotesize\faSquare}} have been fitted to the data. b) Marginal effect of the mean indegree on the relationship between team size and productivity (NLOC). c) Increase of mean indegree (InD) with team size.}
    \label{fig:teamsize_productivity}
\end{figure*}

We now address the ongoing scientific discourse on the relationship between team size and productivity \cite{sornette2014much, scholtes2014causality, muric2019collaboration}. 
For this, we apply linear and polynomial regression models to different productivity target variables, where we additionally use the network metrics identified in section \Cref{sec:colaboration_nets} as control variables.

\begin{table}[p!]
\caption{Regression models relating team size (TS) to five productivity measures. The results are based on 1,188 observations. All productivity measures are log-transformed. All coefficients are Bonferroni-corrected for multiple hypotheses testing.}
\begin{center}
\begin{small}\sffamily

a) Linear relationship\\[1mm]
\begin{tabular}{@{}l@{} D{.}{.}{4.5}@{\hspace*{-2mm}} D{.}{.}{4.5}@{\hspace*{-2mm}} D{.}{.}{4.5}@{\hspace*{-2mm}} D{.}{.}{4.5}@{\hspace*{-2mm}} D{.}{.}{4.5}@{\hspace*{-2mm}} D{.}{.}{4.5}@{\hspace*{-2mm}} D{.}{.}{4.5}@{\hspace*{-2mm}} D{.}{.}{4.5}@{}}
\toprule
 & \multicolumn{1}{c}{Comms} & \multicolumn{1}{c}{Events} & \multicolumn{1}{c}{LevD} & \multicolumn{1}{c}{CycC} & \multicolumn{1}{c}{NLOC} & \multicolumn{1}{c}{Tokens} & \multicolumn{1}{c}{Funcs} & \multicolumn{1}{c}{HalEff} \\
\midrule
(IC)       & 3.38^{***}  & 7.75^{***}  & 11.24^{***} & 5.22^{***}  & 6.99^{***}  & 9.02^{***}  & 4.22^{***}  & 15.82^{***} \\
           & (0.08)      & (0.10)      & (0.10)      & (0.11)      & (0.10)      & (0.11)      & (0.11)      & (0.16)      \\
TS (log)   & -0.20^{***} & -0.31^{***} & -0.33^{***} & -0.31^{***} & -0.31^{***} & -0.34^{***} & -0.29^{***} & -0.22^{***} \\
\phantom{TS$^2$ (log)}           & (0.02)      & (0.03)      & (0.03)      & (0.03)      & (0.03)      & (0.03)      & (0.03)      & (0.04)      \\
\midrule
R$^2$      & 0.07        & 0.10        & 0.11        & 0.09        & 0.10        & 0.11        & 0.08        & 0.02        \\
Adj. R$^2$ & 0.07        & 0.10        & 0.11        & 0.09        & 0.10        & 0.11        & 0.08        & 0.02        \\
% Num. obs.  & 1188        & 1188        & 1188        & 1188        & 1188        & 1188        & 1188        & 1188        \\
\bottomrule
\end{tabular}
% \begin{tabular}{@{}l@{} D{.}{.}{4.5}@{} D{.}{.}{4.5}@{} D{.}{.}{4.5}@{} D{.}{.}{4.5}@{} D{.}{.}{4.5}@{}}
% \toprule
%  & \multicolumn{1}{c}{Commits} & \multicolumn{1}{c}{LevD} & \multicolumn{1}{c}{CycC} & \multicolumn{1}{c}{NLOC} & \multicolumn{1}{c}{HalEff} \\
% \midrule
% (IC)         & 3.38^{***}  & 11.24^{***} & 5.22^{***}  & 6.99^{***}  & 15.82^{***} \\
%              & (0.08)      & (0.10)      & (0.11)      & (0.10)      & (0.16)      \\
% TS (log)     & -0.20^{***} & -0.33^{***} & -0.31^{***} & -0.31^{***} & -0.22^{***} \\
%              & (0.02)      & (0.03)      & (0.03)      & (0.03)      & (0.04)      \\
% \midrule
% R$^2$        & 0.07        & 0.11        & 0.09        & 0.10        & 0.02        \\
% Adj. R$^2$   & 0.07        & 0.11        & 0.09        & 0.10        & 0.02        \\
% % Num. obs.  & 1188        & 1188        & 1188        & 1188        & 1188        \\
% \bottomrule
% \end{tabular}
\\
\vspace{2.5mm}
b) Quadratic relationship\\[1mm]
\begin{tabular}{@{}l@{} D{.}{.}{4.5}@{\hspace*{-2mm}} D{.}{.}{4.5}@{\hspace*{-2mm}} D{.}{.}{4.5}@{\hspace*{-2mm}} D{.}{.}{4.5}@{\hspace*{-2mm}} D{.}{.}{4.5}@{\hspace*{-2mm}} D{.}{.}{4.5}@{\hspace*{-2mm}} D{.}{.}{4.5}@{\hspace*{-2mm}} D{.}{.}{4.5}@{}}
\toprule
 & \multicolumn{1}{c}{Comms} & \multicolumn{1}{c}{Events} & \multicolumn{1}{c}{LevD} & \multicolumn{1}{c}{CycC} & \multicolumn{1}{c}{NLOC} & \multicolumn{1}{c}{Tokens} & \multicolumn{1}{c}{Funcs} & \multicolumn{1}{c}{HalEff} \\
\midrule
(IC)         & 3.02^{***} & 7.06^{***}  & 10.55^{***} & 4.18^{***}  & 6.23^{***}  & 8.11^{***}  & 3.31^{***}  & 14.10^{***} \\
             & (0.17)     & (0.22)      & (0.22)      & (0.23)      & (0.21)      & (0.22)      & (0.23)      & (0.34)      \\
TS (log)     & 0.02       & 0.12        & 0.10        & 0.35^{*}   & 0.17        & 0.24        & 0.28    & 0.86^{***}  \\
             & (0.10)     & (0.12)      & (0.13)      & (0.13)      & (0.12)      & (0.13)      & (0.13)      & (0.19)      \\
TS$^2$ (log) & -0.03^{*}  & -0.06^{***} & -0.06^{***} & -0.09^{***} & -0.06^{***} & -0.08^{***} & -0.08^{***} & -0.15^{***} \\
             & (0.01)     & (0.02)      & (0.02)      & (0.02)      & (0.02)      & (0.02)      & (0.02)      & (0.03)      \\
\midrule
R$^2$        & 0.08       & 0.11        & 0.12        & 0.11        & 0.11        & 0.13        & 0.10        & 0.05        \\
Adj. R$^2$   & 0.07       & 0.11        & 0.12        & 0.11        & 0.11        & 0.12        & 0.10        & 0.05        \\
% Num. obs.    & 1188       & 1188        & 1188        & 1188        & 1188        & 1188        & 1188        & 1188        \\
\bottomrule
\end{tabular}
% \begin{tabular}{@{}l@{} D{.}{.}{4.5}@{} D{.}{.}{4.5}@{} D{.}{.}{4.5}@{} D{.}{.}{4.5}@{} D{.}{.}{4.5}@{}}
% \toprule
%  & \multicolumn{1}{c}{Commits} & \multicolumn{1}{c}{LevD} & \multicolumn{1}{c}{CycC} & \multicolumn{1}{c}{NLOC} & \multicolumn{1}{c}{HalEff} \\
% \midrule
% (IC)         & 3.02^{***} & 10.55^{***} & 4.18^{***}  & 6.23^{***}  & 14.10^{***} \\
%              & (0.17)     & (0.22)      & (0.23)      & (0.21)      & (0.34)      \\
% TS (log)     & 0.02       & 0.10        & 0.35^{*}    & 0.17        & 0.86^{***}  \\
%              & (0.10)     & (0.13)      & (0.13)      & (0.12)      & (0.19)      \\
% TS$^2$ (log) & -0.03^{*}  & -0.06^{***} & -0.09^{***} & -0.06^{***} & -0.15^{***} \\
%              & (0.01)     & (0.02)      & (0.02)      & (0.02)      & (0.03)      \\
% \midrule
% R$^2$        & 0.08       & 0.12        & 0.11        & 0.11        & 0.05        \\
% Adj. R$^2$   & 0.07       & 0.12        & 0.11        & 0.11        & 0.05        \\
% % Num. obs.  & 1188       & 1188        & 1188        & 1188        & 1188        \\
% \bottomrule
% \end{tabular}
\\
\vspace{2.5mm}
c) Linear relationship controlling for network properties\\[1mm]
\begin{tabular}{@{}l@{} D{.}{.}{4.5}@{\hspace*{-2mm}} D{.}{.}{4.5}@{\hspace*{-2mm}} D{.}{.}{4.5}@{\hspace*{-2mm}} D{.}{.}{4.5}@{\hspace*{-2mm}} D{.}{.}{4.5}@{\hspace*{-2mm}} D{.}{.}{4.5}@{\hspace*{-2mm}} D{.}{.}{4.5}@{\hspace*{-2mm}} D{.}{.}{4.5}@{}}
\toprule
 & \multicolumn{1}{c}{Comms} & \multicolumn{1}{c}{Events} & \multicolumn{1}{c}{LevD} & \multicolumn{1}{c}{CycC} & \multicolumn{1}{c}{NLOC} & \multicolumn{1}{c}{Tokens} & \multicolumn{1}{c}{Funcs} & \multicolumn{1}{c}{HalEff} \\
\midrule
(IC)       & 3.18^{***}  & 7.62^{***}  & 11.10^{***} & 5.22^{***}  & 6.95^{***}  & 8.93^{***}  & 4.25^{***}  & 15.78^{***} \\
           & (0.08)      & (0.09)      & (0.10)      & (0.10)      & (0.09)      & (0.10)      & (0.10)      & (0.16)      \\
TS (log)   & -0.36^{***} & -0.49^{***} & -0.51^{***} & -0.48^{***} & -0.48^{***} & -0.52^{***} & -0.45^{***} & -0.45^{***} \\
           & (0.02)      & (0.02)      & (0.02)      & (0.02)      & (0.02)      & (0.02)      & (0.03)      & (0.04)      \\
InD (log)  & 1.16^{***}  & 1.45^{***}  & 1.44^{***}  & 1.46^{***}  & 1.42^{***}  & 1.49^{***}  & 1.35^{***}  & 1.90^{***}  \\
           & (0.04)      & (0.05)      & (0.06)      & (0.06)      & (0.05)      & (0.06)      & (0.06)      & (0.09)      \\
FModR      & -1.00^{***} & -1.90^{***} & -1.83^{***} & -2.65^{***} & -2.33^{***} & -2.19^{***} & -2.65^{***} & -3.26^{***} \\
           & (0.22)      & (0.27)      & (0.28)      & (0.30)      & (0.27)      & (0.28)      & (0.30)      & (0.46)      \\
\midrule
R$^2$      & 0.45        & 0.49        & 0.47        & 0.46        & 0.49        & 0.49        & 0.42        & 0.33        \\
Adj. R$^2$ & 0.45        & 0.49        & 0.46        & 0.46        & 0.49        & 0.49        & 0.42        & 0.33        \\
% Num. obs.  & 1188        & 1188        & 1188        & 1188        & 1188        & 1188        & 1188        & 1188        \\
\bottomrule
\end{tabular}
% \begin{tabular}{@{}l@{} D{.}{.}{4.5}@{} D{.}{.}{4.5}@{} D{.}{.}{4.5}@{} D{.}{.}{4.5}@{} D{.}{.}{4.5}@{}}
% \toprule
%  & \multicolumn{1}{c}{Commits} & \multicolumn{1}{c}{LevD} & \multicolumn{1}{c}{CycC} & \multicolumn{1}{c}{NLOC} & \multicolumn{1}{c}{HalEff} \\
% \midrule
% (IC)         & 3.18^{***}  & 11.10^{***} & 5.22^{***}  & 6.95^{***}  & 15.78^{***} \\
%              & (0.08)      & (0.10)      & (0.10)      & (0.09)      & (0.16)      \\
% TS (log)     & -0.36^{***} & -0.51^{***} & -0.48^{***} & -0.48^{***} & -0.45^{***} \\
%              & (0.02)      & (0.02)      & (0.02)      & (0.02)      & (0.04)      \\
% InD (log)    & 1.16^{***}  & 1.44^{***}  & 1.46^{***}  & 1.42^{***}  & 1.90^{***}  \\
%              & (0.04)      & (0.06)      & (0.06)      & (0.05)      & (0.09)      \\
% FModR        & -1.00^{***} & -1.83^{***} & -2.65^{***} & -2.33^{***} & -3.26^{***} \\
%              & (0.22)      & (0.28)      & (0.30)      & (0.27)      & (0.46)      \\
% \midrule
% R$^2$        & 0.45        & 0.47        & 0.46        & 0.49        & 0.33        \\
% Adj. R$^2$   & 0.45        & 0.46        & 0.46        & 0.49        & 0.33        \\
% % Num. obs.  & 1188        & 1188        & 1188        & 1188        & 1188        \\
% \bottomrule
% \end{tabular}
\end{small}
\label{tab:reg_models}
\end{center}
\end{table}

\addtocounter{table}{-1}

\begin{table}[p!]
\caption{(continued) Regression models relating team size (TS) to five productivity measures. The results are based on 1,188 observations. All productivity measures are log-transformed. All coefficients are Bonferroni-corrected for multiple hypotheses testing.}
\begin{center}
\begin{small}\sffamily
d) Quadratic relationship controlling for network properties\\[1mm]
\begin{tabular}{@{}l@{} D{.}{.}{4.5}@{\hspace*{-2mm}} D{.}{.}{4.5}@{\hspace*{-2mm}} D{.}{.}{4.5}@{\hspace*{-2mm}} D{.}{.}{4.5}@{\hspace*{-2mm}} D{.}{.}{4.5}@{\hspace*{-2mm}} D{.}{.}{4.5}@{\hspace*{-2mm}} D{.}{.}{4.5}@{\hspace*{-2mm}} D{.}{.}{4.5}@{}}
\toprule
 & \multicolumn{1}{c}{Comms} & \multicolumn{1}{c}{Events} & \multicolumn{1}{c}{LevD} & \multicolumn{1}{c}{CycC} & \multicolumn{1}{c}{NLOC} & \multicolumn{1}{c}{Tokens} & \multicolumn{1}{c}{Funcs} & \multicolumn{1}{c}{HalEff} \\
\midrule
(IC)         & 3.53^{***}  & 7.83^{***}  & 11.30^{***} & 5.10^{***}  & 7.08^{***}  & 8.95^{***}  & 4.22^{***}  & 15.25^{***} \\
             & (0.14)      & (0.17)      & (0.18)      & (0.19)      & (0.17)      & (0.18)      & (0.19)      & (0.30)      \\
TS (log)     & -0.58^{***} & -0.63^{***} & -0.64^{***} & -0.41^{***} & -0.56^{***} & -0.53^{***} & -0.42^{***} & -0.10       \\
             & (0.08)      & (0.10)      & (0.10)      & (0.11)      & (0.10)      & (0.10)      & (0.11)      & (0.17)      \\
TS$^2$ (log) & 0.03^{*}   & 0.02        & 0.02        & -0.01       & 0.01        & 0.00        & -0.00       & -0.05   \\
             & (0.01)      & (0.01)      & (0.01)      & (0.01)      & (0.01)      & (0.01)      & (0.01)      & (0.02)      \\
InD (log)    & 1.18^{***}  & 1.46^{***}  & 1.45^{***}  & 1.45^{***}  & 1.43^{***}  & 1.49^{***}  & 1.35^{***}  & 1.87^{***}  \\
             & (0.04)      & (0.05)      & (0.06)      & (0.06)      & (0.05)      & (0.06)      & (0.06)      & (0.09)      \\
FModR        & -1.04^{***} & -1.93^{***} & -1.85^{***} & -2.64^{***} & -2.34^{***} & -2.19^{***} & -2.65^{***} & -3.20^{***} \\
             & (0.22)      & (0.27)      & (0.28)      & (0.30)      & (0.27)      & (0.28)      & (0.30)      & (0.46)      \\
\midrule
R$^2$        & 0.45        & 0.49        & 0.47        & 0.46        & 0.49        & 0.49        & 0.42        & 0.34        \\
Adj. R$^2$   & 0.45        & 0.49        & 0.46        & 0.46        & 0.49        & 0.49        & 0.42        & 0.33        \\
% Num. obs.    & 1188        & 1188        & 1188        & 1188        & 1188        & 1188        & 1188        & 1188        \\
\bottomrule
\end{tabular}
% \begin{tabular}{@{}l@{} D{.}{.}{4.5}@{} D{.}{.}{4.5}@{} D{.}{.}{4.5}@{} D{.}{.}{4.5}@{} D{.}{.}{4.5}@{}}
% \toprule
%  & \multicolumn{1}{c}{Commits} & \multicolumn{1}{c}{LevD} & \multicolumn{1}{c}{CycC} & \multicolumn{1}{c}{NLOC} & \multicolumn{1}{c}{HalEff} \\
% \midrule
% (IC)           & 3.53^{***}  & 11.30^{***} & 5.10^{***}  & 7.08^{***}  & 15.25^{***} \\
%               & (0.14)      & (0.18)      & (0.19)      & (0.17)      & (0.30)      \\
% TS (log)       & -0.58^{***} & -0.64^{***} & -0.41^{***} & -0.56^{***} & -0.10       \\
%               & (0.08)      & (0.10)      & (0.11)      & (0.10)      & (0.17)      \\
% TS$^2$ (log)   & 0.03^{*}    & 0.02        & -0.01       & 0.01        & -0.05   \\
%               & (0.01)      & (0.01)      & (0.01)      & (0.01)      & (0.02)      \\
% InD (log)      & 1.18^{***}  & 1.45^{***}  & 1.45^{***}  & 1.43^{***}  & 1.87^{***}  \\
%               & (0.04)      & (0.06)      & (0.06)      & (0.05)      & (0.09)      \\
% FModR          & -1.04^{***} & -1.85^{***} & -2.64^{***} & -2.34^{***} & -3.20^{***} \\
%               & (0.22)      & (0.28)      & (0.30)      & (0.27)      & (0.46)      \\
% \midrule
% R$^2$          & 0.45        & 0.47        & 0.46        & 0.49        & 0.34        \\
% Adj. R$^2$     & 0.45        & 0.46        & 0.46        & 0.49        & 0.33        \\
% % Num. obs.    & 1188        & 1188        & 1188        & 1188        & 1188        \\
% \bottomrule
% \end{tabular}
\\
\vspace{2.5mm}
e) Linear relationship with controls and interaction effects\\[1mm]
\begin{tabular}{@{}l@{} D{.}{.}{4.5}@{\hspace*{-2mm}} D{.}{.}{4.5}@{\hspace*{-2mm}} D{.}{.}{4.5}@{\hspace*{-2mm}} D{.}{.}{4.5}@{\hspace*{-2mm}} D{.}{.}{4.5}@{\hspace*{-2mm}} D{.}{.}{4.5}@{\hspace*{-2mm}} D{.}{.}{4.5}@{\hspace*{-2mm}} D{.}{.}{4.5}@{}}
\toprule
 & \multicolumn{1}{c}{Comms} & \multicolumn{1}{c}{Events} & \multicolumn{1}{c}{LevD} & \multicolumn{1}{c}{CycC} & \multicolumn{1}{c}{NLOC} & \multicolumn{1}{c}{Tokens} & \multicolumn{1}{c}{Funcs} & \multicolumn{1}{c}{HalEff} \\
\midrule
(IC)          & 2.72^{***}  & 7.39^{***}  & 10.85^{***} & 5.01^{***}  & 6.68^{***}  & 8.68^{***}  & 4.06^{***}  & 15.91^{***} \\
              & (0.11)      & (0.14)      & (0.14)      & (0.15)      & (0.13)      & (0.14)      & (0.15)      & (0.23)      \\
TS (log)      & -0.22^{***} & -0.43^{***} & -0.44^{***} & -0.42^{***} & -0.40^{***} & -0.45^{***} & -0.39^{***} & -0.48^{***} \\
              & (0.03)      & (0.04)      & (0.04)      & (0.04)      & (0.04)      & (0.04)      & (0.04)      & (0.06)      \\
InD (log)     & 1.83^{***}  & 1.78^{***}  & 1.80^{***}  & 1.76^{***}  & 1.80^{***}  & 1.86^{***}  & 1.62^{***}  & 1.72^{***}  \\
              & (0.12)      & (0.15)      & (0.16)      & (0.17)      & (0.15)      & (0.16)      & (0.17)      & (0.26)      \\
TS$\times$InD & -0.18^{***} & -0.09   & -0.10   & -0.08       & -0.10^{*}  & -0.10   & -0.07       & 0.05        \\
              & (0.03)      & (0.04)      & (0.04)      & (0.04)      & (0.04)      & (0.04)      & (0.04)      & (0.07)      \\
FModR         & -0.92^{***} & -1.87^{***} & -1.79^{***} & -2.62^{***} & -2.29^{***} & -2.15^{***} & -2.62^{***} & -3.28^{***} \\
              & (0.22)      & (0.27)      & (0.28)      & (0.30)      & (0.27)      & (0.28)      & (0.30)      & (0.46)      \\
\midrule
R$^2$         & 0.47        & 0.49        & 0.47        & 0.46        & 0.50        & 0.49        & 0.43        & 0.34        \\
Adj. R$^2$    & 0.46        & 0.49        & 0.47        & 0.46        & 0.49        & 0.49        & 0.42        & 0.33        \\
% Num. obs.     & 1188        & 1188        & 1188        & 1188        & 1188        & 1188        & 1188        & 1188        \\
\bottomrule
\multicolumn{9}{l}{\vphantom{\Large A}$^{***}p<0.001$; $^{**}p<0.01$; $^{*}p<0.05$}
\end{tabular}
% \begin{tabular}{@{}l@{} D{.}{.}{4.5}@{} D{.}{.}{4.5}@{} D{.}{.}{4.5}@{} D{.}{.}{4.5}@{} D{.}{.}{4.5}@{}}
% \toprule
%  & \multicolumn{1}{c}{Commits} & \multicolumn{1}{c}{LevD} & \multicolumn{1}{c}{CycC} & \multicolumn{1}{c}{NLOC} & \multicolumn{1}{c}{HalEff} \\
% \midrule
% (IC)            & 2.72^{***}  & 10.85^{***} & 5.01^{***}  & 6.68^{***}  & 15.91^{***} \\
%                 & (0.11)      & (0.14)      & (0.15)      & (0.13)      & (0.23)      \\
% TS (log)        & -0.22^{***} & -0.44^{***} & -0.42^{***} & -0.40^{***} & -0.48^{***} \\
%                 & (0.03)      & (0.04)      & (0.04)      & (0.04)      & (0.06)      \\
% InD (log)       & 1.83^{***}  & 1.80^{***}  & 1.76^{***}  & 1.80^{***}  & 1.72^{***}  \\
%                 & (0.12)      & (0.16)      & (0.17)      & (0.15)      & (0.26)      \\
% TS$\times$InD   & -0.18^{***} & -0.10       & -0.08       & -0.10^{*}   & 0.05        \\
%                 & (0.03)      & (0.04)      & (0.04)      & (0.04)      & (0.07)      \\
% FModR           & -0.92^{***} & -1.79^{***} & -2.62^{***} & -2.29^{***} & -3.28^{***} \\
%                 & (0.22)      & (0.28)      & (0.30)      & (0.27)      & (0.46)      \\
% \midrule
% R$^2$           & 0.47        & 0.47        & 0.46        & 0.50        & 0.34        \\
% Adj. R$^2$      & 0.46        & 0.47        & 0.46        & 0.49        & 0.33        \\
% % Num. obs.     & 1188        & 1188        & 1188        & 1188        & 1188        \\
% \bottomrule
% \multicolumn{6}{l}{\vphantom{\Large a}$^{***}p<0.001$; $^{**}p<0.01$; $^{*}p<0.05$}
% \end{tabular}
\end{small}
%\label{tab:reg_models}
\end{center}
\end{table}

\Cref{fig:teamsize_productivity}a shows the productivity per team member as a function of team size, exemplified for a productivity measure based on cyclomatic complexity.
We fit a linear ($\text{CycC} \sim \text{TS}$) and a polynomial model with maximum degree two ($\text{CycC} \sim \text{TS} + \text{TS}^2$) to our data.
The linear model enables us to infer a possible relationship between team size and productivity.
The model $\text{CycC} \sim \text{TS} + \text{TS}^2$ is the basis to test for the existence of an optimal team size, which is captured by the existence of a global maximum of the quadratic function.

The linear model yields a significant negative relationship between TS and productivity, as reported in \Cref{tab:reg_models}a.
Similarly, as shown in \Cref{tab:reg_models}b, we find a significant negative coefficient for TS$^2$ for the quadratic model.
This means that, on average, individual productivity decreases with team size.

That said, all coefficients for TS in the quadratic models are positive, and the coefficients for CycC and HalEff are also significant.
This means that these models can be represented as inverted parabolas, e.g., as shown by the red curve for CycC in \Cref{fig:teamsize_productivity}a.
The maxima of the parabolas for CycC and HalEff provide some evidence for an optimal team size of 7 or 19 team members, respectively, which is roughly in line with the optimal team size of 9 suggested by \citep{rodriguez2012empirical}.
However, we argue that the key insight of this result is not the \emph{exact} team size for which the maximum is reached, which is likely an artefact of the simplified model used for our analysis.
Instead, the key insight is the possible increase in individual productivity for very small teams---compared to the analysed range from 2 to 1,711 members---with decreases thereafter.
Due to the large amount of remaining variance and the small slope of the parabola around the maximum, interpreting a specific number as optimal team size is likely to be misleading.
We further note that for the regression analyses using the remaining productivity measures as target variables, the coefficient for team size is insignificant.
Hence, despite the coefficients being consistently positive, these models do not provide sufficient statistical evidence to conclude an increase in individual productivity even in small teams, resulting in an overall negative relation between productivity and team size.

Importantly, the models considered so far only partially explain the variance in the relationship between productivity and team size.
This can be seen from the low $R^2$ values of less than 15\% in \Cref{tab:reg_models}a and b.
Moreover, \Cref{fig:teamsize_productivity}a shows that, especially for small teams, the productivity of team members varies over four decades.
This suggests that additional aspects other than team size have a substantial influence on the productivity of developers.
We thus consider regression models that additionally control for the network features identified in \Cref{sec:colaboration_nets}.
Specifically, we add the mean number of interaction partners per team member (InD) and the average amount of edited foreign code (FModR) as additional variables in our regression models.

As shown by the $R^2$ values in \Cref{tab:reg_models}c, the regression models that include InD and FModR explain close to half of the variance in the productivity observations.
We further again find a negative relationship between team size and productivity for all five operationalisations.
Assuming constant InD and FModR, the regression results suggest that by doubling the size of a development team, we reduce the average productivity of team members by between 22\% and 30\%.
Notably, a higher mean indegree is accompanied by higher productivity.
Moreover, the foreign modification ratio has a strong negative relationship with productivity.

Adding a quadratic term ($\text{TS}^2$) to the model, i.e.,
\begin{align}
    \text{PROD} \sim \text{TS} + \text{TS}^2 + \text{InD} + \text{FModR},
\end{align}
we find that contrary to before, the coefficient of team size remains negative and significant while team size squared is insignificant (see \Cref{tab:reg_models}d).
Thus, the non-linear relationship between productivity and team size found in \Cref{tab:reg_models}b does not persist when accounting for the mean indegree and foreign modification ratio.

In conclusion, in a large-scale study using 201 collaborative \textit{GitHub} projects sampled in a systematic and unbiased fashion across different strata of team sizes, we confirm the negative relationship between team size and productivity found by prior studies.
This negative relationship is robust against the choice of productivity measure and persists when controlling for the team's collaboration structure.
Only considering team size as a predictor, our data provide some evidence for an optimal team size of 7 or 19 members for two of the eight operationalisations of productivity.
However, for six out of eight productivity measures, the optimal productivity per team member is reached for a team size of one.
This finding is further supported by the fact that the squared team size has no significant relationship when controlling for other network measures.

As mentioned above, our results suggest that the mean indegree of developers is positively related to productivity, while team size is negatively related to productivity.
Importantly, the regression models considered so far assume that the effect of the team size on productivity is independent of the mean indegree and vice-versa, i.e., their effect is purely additive.
However, it is reasonable to assume that the productivity of developers in a team with dense collaboration structures is more strongly affected by team size, compared to a team where each developer collaborates, on average, with few other team members.
This motivates a last experiment, where we include an interaction term that captures the combined effect of team size and mean indegree as an additional variable:
\begin{align}
    \text{PROD} \sim \text{TS} + \text{InD} + \text{TS}\times\text{InD} + \text{FModR}
\end{align}
The results in \Cref{tab:reg_models}e suggest a negative coefficient for this interaction term.
However, the effect is only significant for Comms and NLOC.
The analysis of the marginal effect of the mean indegree on the relationship between team size and productivity is shown in \Cref{fig:teamsize_productivity}b.
In line with the positive coefficient of the mean indegree, we find that developers in teams with larger mean indegree tend to be more productive on average, i.e. lines corresponding to larger mean indegrees tend to have larger intercepts (but negative slopes).
Moreover, as shown by the negative coefficient of the interaction term, the negative effect of team size on productivity grows with the mean indegree, i.e. lines corresponding to larger mean indegrees tend to have steeper negative slopes.
The whisker plot in \Cref{fig:teamsize_productivity}c further reveals a positive relationship between the size of a team (x-axis) and the mean indegree of developers (y-axis), i.e. developers in larger teams tend to edit code of a larger number of other developers.
This positive relationship specifically holds for smaller team sizes, while the mean indegree in larger teams with more than approximately 50 developers is similar.
Importantly, the fact that (i) developers in larger teams tend to have a higher indegree and (ii) developers in teams with higher indegree tend to be more productive does \emph{not} imply that developers in larger teams are, on average, more productive.
This is confirmed by the negative coefficients of team size in all our regression models as well as the clear negative marginal effects shown in \Cref{fig:teamsize_productivity}b.

We note that the positive relationship between team size and the mean indegree could explain the positive coefficient for TS in \Cref{tab:reg_models}b that suggests the existence of an optimal team size.
We further conjecture that this non-trivial finding could be a reason why empirical studies that do not account for the distribution of team sizes in \textit{GitHub} repositories, and thus inadvertently focus on projects with small team size, erroneously find a positive relationship between team size and productivity.
Specifically, projects with small team sizes tend to have a small mean indegree, which corresponds to a line with smaller intercept (and negative slope) in \Cref{tab:reg_models}b.
Conversely, teams with larger team sizes tend to have a larger mean indegree, which corresponds to a line with larger intercept (and negative slope) in \Cref{tab:reg_models}b.
The failure to control for this effect can lead to a reversal of the relationship between productivity and team size, i.e., a wrong \emph{positive} slope for the effect is erroneously found.
This can be viewed as a specific instance of Simpson's paradox, where the aggregate effect in a sample that combines data from different ``groups'' of projects---i.e., teams with different mean indegrees---can be positive, even though a negative relationship holds for each group separately.

\FloatBarrier

\section{Limitations and Threats to Validity}

A first threat to the validity of our results could be the operationalisation of productivity.
To guard against this, we have studied eight different measures that capture different notions of productivity.
A common issue of productivity measures that are based on commit log data is that they do not account for the structure of contributed code.
To guard against this issue and appropriately value commits that decrease the complexity of code and thus make it more maintainable, we consider measures that account for tokens, functions, and control structures. 

A second aspect that could potentially influence our results is the method to assess the size of a software team.
We compute the team size at a given time $t$ by counting all developers who have made a commit up to $42$ weeks before $t$.
This approach to infer the team size is necessary since there is no formalised notion of team size in \textit{GitHub}.
The specific choice of this time window is based on the inter-commit time distribution for \textit{GitHub} projects found in \citep{scholtes2016aristotle}.
We have tested the robustness of the results by choosing a different window size of two years. 
Due to the computational effort that is due to the recalculation of all network metrics, a comprehensive study of different window sizes was beyond the scope of our study but could be an interesting question for future work. 

In \Cref{sec:challenge_of_selecting_software_repositories}, we identified project selection and data preparation as a major threat to the validity.
We thus spent considerable effort to develop a general project selection pipeline as well as Open Source software tools to infer collaboration networks from commit data.
Despite these efforts, there may be remaining issues, such as the possibility to manually modify the history of a \textit{git} repository, which we can neither detect nor account for.
Due to data issues related to some merge commits, we  were further not able to process all commits of the \texttt{Linux Kernel} project\footnote{\url{https://github.com/torvalds/linux}}.
We therefore excluded this project from our analysis.

Addressing the issue of omitted variables, our regression models explain roughly 50\% of the variance in the relationship between team size and productivity, which considerably improves the variance explained by prior studies.
Nevertheless, there is additional variance in the relationship that we cannot explain.
This could either be due to the stochastic nature of the underlying process or the existence of additional variables that are not included in our models.
To address this issue, future studies could additionally study data from issue trackers and mailing lists \citep{bacchelli2011miler, bird2006mining, guzzi2013communication}.

In our study, we considered 201 OSS projects sampled to represent the entire spectrum of team sizes present on \textit{GitHub}.
While our findings are representative for collaborative software development on \textit{GitHub}, it is unclear whether they can be generalised to proprietary software projects or other Open Source collaboration platforms.
Platforms like SourceForge or Bitbucket have different characteristics \citep{howison2004perils, ma2019world, xie2009data} that will require modifications to our selection pipeline, which is an interesting issue for future work.

\section{Related Work}\label{sec:related_work}

Our work touches on issues that have been studied in empirical software engineering, computational social science, network science, and organisational theory. Namely, how we can use repository log data to quantify productivity in software development, how the size of teams influences the productivity of its members, and how network models can be used to study social aspects in collaborative software projects.
At a meta-level, our study further addresses common challenges and pitfalls in the analysis of big repository data, some of which have been previously highlighted in \citep{bird2009promises,kalliamvakou2014promises,howison2004perils,kalliamvakou2016depth}.

A large body of works has investigated methods to measure the \emph{productivity} of developers based on repository data \cite{meyer2014software, sudhakar2012measuring}.
\emph{Commit-based productivity measures} calculate productivity based on the number of commits \cite{mockus2002two, sornette2014much} or pushes \cite{muric2019collaboration}.
While this approach does not require a detailed analysis of the committed source code, it has the problem that the amount of code changed with each commit can vary significantly both within and between projects \cite{github2020weeklypattern}.
\cite{oliveira2020code} found that productivity rankings based on commit-based measures only show low correlations with rankings obtained from team leaders. 
Therefore, more fine-granular measures such as the number of modified lines or the number of modified characters \cite{scholtes2016aristotle} have been proposed.
\emph{Code-based productivity measures} aim to overcome this limitation by analysing the code contained in a commit.
Commonly used metric include the number of code lines \cite{devanbu1996analytical, blackburn1996improving, hulkko2005multiple, nguyen2011analysis}, function points \cite{jones1994software, wagner2018systematic}, or tokens \cite{lane1997intergrating} changed per time interval.

A large body of works in organisational theory, computational social science and empirical software engineering have studied the question of how the size of a team is related to its performance. 
In the context of software development, \cite{brooks1975mythical} argues that the increased coordination requirements in larger teams lead to a reduction of developer productivity.
While this proposed mechanism has been corroborated by quantitative studies \cite{jiang2007investigation, scholtes2016aristotle}, other works suggest synergistic effects that lead to an increase in developer productivity as teams grow in size \cite{sornette2014much, muric2019collaboration}.
The combination of these findings indicates that an optimal size for a software development team may exist, which is also discussed in the literature \cite{herivcko2008approach, rodriguez2012empirical}.
A report cited by \cite{rodriguez2012empirical} suggests that for proprietary software development projects, an optimal team size with respect to productivity is achieved for nine team members.

Utilising a method to construct a network representation of co-editing relations from the commit-log history of a project, our work finally addresses questions that received attention from the network science and computational social science community.
A number of works have investigated how the topology of communication, collaboration, or coordination networks is related to the performance \cite{wu2013social,yang2004team,reagans2001networks}, resilience \cite{massari2021team,zanetti2013rise}, or productivity \cite{scholtes2016aristotle} of teams in various contexts.
In the context of research on developer productivity, recent studies found that a densification of co-editing networks due to shared code ownership can explain the decrease in productivity observed for larger teams \cite{scholtes2016aristotle,gote2019analysing}.

\section{Conclusion}

Massive data from software repositories and collaboration tools provide compelling new opportunities to study social aspects in software development.
Within this context, the question of how the size and collaboration patterns of software development teams influence the productivity of developers has emerged as an important research question at the intersection of computational social science and empirical software engineering.
Recent empirical studies using big data from software repositories have come to contradictory answers to this important research question, even though those studies used similar data sets and empirical methods.

Addressing common challenges and pitfalls in the analysis of big repository data, our work offers a possible explanation for this disagreement between recent works in empirical software engineering.
To this end, we provide the, to the best of our knowledge, largest, curated corpus of \textit{GitHub} projects that is specifically tailored to investigate the influence of team size and collaboration patterns on individual and collective productivity.
The projects included in this corpus are \emph{systematically} chosen such that we avoid common perils in \textit{GitHub} mining. 
We use a stratified sampling that supports unbiased analyses of the impact of team size on developer productivity.
We systematically compare a set of eight code- and commit-based productivity measures and study which of the measures are likely to be interchangeable and which capture independent dimensions of productivity.
Building on a method to construct time-evolving co-editing networks from \textit{git} repositories, we consider eight network metrics that capture different dimensions of the social organisation of software teams.
We finally use those methods to study the Ringelmann effect in collaborative software development. 
Our results highlight a robust negative relationship between team size and developer productivity that can be explained based on the team's collaboration structure.
We argue that neglecting the highly skewed distribution of team sizes on \textit{GitHub} can lead to a reversal of the relationship between team size and productivity, thus offering a possible explanation for recent contradictory results.

Apart from this, our work provides several insights that are relevant for the management of software projects:
In particular, we find (i) an overall negative relation between individual productivity and team size, and (ii) a non-linear relationship that gives rise to an optimal team size for small teams.
These findings can be useful to define advanced cost estimation models that incorporate the found non-linear relationship between team size and productivity, thus providing better estimates for the work force required to develop projects with a known (estimated) size of the code base.
Our analysis further highlights additional factors that influence team productivity, such as the amount of foreign code that is edited and the number of interaction partners of developers.
This insight not only allows us to further improve cost estimation models, it also points to factors that can possibly be optimised by project maintainers, e.g., by carefully decomposing the code base into modules addressed by different (sub-)teams or by optimising organisational structures of the development team.

In summary, our work contributes to the ongoing discussion on how the size and structure of teams influence productivity.
Investigating the cross-correlations of productivity and network metrics in a systematically constructed corpus of software projects, we further contribute a valuable new resource for researchers in empirical software engineering and computational social science.
By focusing on generic network measures, we further provide the perspective that our results can generalise beyond empirical software engineering.
Highlighting pitfalls in the analysis of big data, our work finally demonstrates that the use of bigger data sets does not automatically lead to more reliable insights.

\section*{Data Availability and Reproducibility}
To facilitate the reproduction of our results and enable future research based on the extensive data set mined for our study, we have archived both a reproducibility package and our full data sets on \url{zenodo.org}\footnote{Reproducibility package: https://doi.org/10.5281/zenodo.5294015, Data sets: https://doi.org/10.5281/zenodo.5294964}.

\begin{acks}
We thank Christian Zingg for contributing to the development of the infrastructure to mine edits and co-edits from software projects on the ETH Zurich scientific compute cluster \textit{Euler}.
Ingo Scholtes acknowledges financial support from the Swiss National Science Foundation through grant no. 176938. Christoph Gote and Ingo Scholtes wrote parts of this manuscript on a joint research retreat at \emph{Niederzerferm\"uhle} that was financially supported by the Department of Informatics at University of Zurich and the Chair of Systems Design at ETH Zurich.
\end{acks}

\end{document}